\documentclass[aps,prb,twocolumn,amsmath,showpacs,amssymb]{revtex4}

\usepackage{graphics}
\usepackage{amsmath}
\usepackage{dcolumn}
\usepackage{citesort}
\usepackage{epsfig}

\begin{document}

\title{{\rm\small\hfill (submitted to Phys. Rev. B)}\\
Non-adiabatic Effects in the Dissociation of Oxygen Molecules at the Al(111) Surface}

\author{J\"org Behler}
\affiliation{Fritz-Haber-Institut der Max-Planck-Gesellschaft,
Faradayweg 4-6, D-14195 Berlin, Germany}

\author{Karsten Reuter}
\affiliation{Fritz-Haber-Institut der Max-Planck-Gesellschaft,
Faradayweg 4-6, D-14195 Berlin, Germany}

\author{Matthias Scheffler}
\affiliation{Fritz-Haber-Institut der Max-Planck-Gesellschaft,
Faradayweg 4-6, D-14195 Berlin, Germany}

\date{\today}

\begin{abstract}
The measured low initial sticking probability of oxygen molecules at
the Al(111) surface that had puzzled the field for many years was
recently explained in a non-adiabatic picture invoking
spin-selection rules [J. Behler {\em et al.}, Phys. Rev. Lett. {\bf
94}, 036104 (2005)]. These selection rules tend to conserve the
initial spin-triplet character of the free O$_2$ molecule during the
molecule's approach to the surface. A new locally-constrained
density-functional theory approach gave access to the corresponding
potential-energy surface (PES) seen by such an impinging
spin-triplet molecule and indicated barriers to dissociation which
reduce the sticking probability. Here, we further substantiate this
non-adiabatic picture by providing a detailed account of the
employed approach. Building on the previous work, we focus in
particular on inaccuracies in present-day exchange-correlation
functionals. Our analysis shows that small quantitative differences
in the spin-triplet constrained PES obtained with different
gradient-corrected functionals have a noticeable effect on the
lowest kinetic energy part of the resulting sticking curve.
\end{abstract}


\maketitle

\section{Introduction}

The interaction of oxygen with metal surfaces plays a central role
in many technologically relevant processes like heterogeneous
catalysis or corrosion. Key elementary steps in this interaction are
the (dissociative) adsorption of the molecule at the surface,
diffusion at or in the surface, as well as the (associative)
desorption from the surface. Aiming to establish an atomic-scale
understanding, detailed investigations study these interaction steps
at well-defined model systems, often employing single crystal
surfaces exposed to defined amounts of oxygen in an otherwise
ultra-high vacuum surrounding. A prominent example are studies of
the dissociative adsorption of oxygen molecules at the Al(111)
surface, which allegedly represents a most simple and basic case:
the initial step in the oxygen interaction with a close-packed
surface of a nearly free electron metal.

Surprisingly, even here such a fundamental issue like the adsorption
mechanism has not yet been settled. Several contradictory models
like the so-called ``hot atom'' motion~\cite{brune92,brune93},
abstraction~\cite{binetti00,komrowski01}, or dissociation leading to
neighboring adsorbed O atoms~\cite{schmid01} have been proposed, mainly based on
different interpretations of scanning tunnelling microscopy (STM)
data. Entangled with the mechanism is the sticking probability,
which is defined as the ratio of sticking (dissociation) events to
the total number of molecule-surface collisions. Many independent
experiments have unambiguously shown that the initial sticking
probability of thermal oxygen molecules at Al(111) is only about
1~\%.\cite{gartland77,osterlund97,bradshaw77,brune93,zhukov99a,lee01}
Furthermore, using molecular beam experiments it was found that the
sticking probability increases with translational kinetic energy of
the impinging molecules and reaches a saturation value of about
90~\% only for kinetic energies higher than
0.5~eV.~\cite{osterlund97} A straightforward explanation for this
finding would be the existence of energy barriers towards
dissociation that cannot be overcome by low energy, thermal
molecules. Several first-principles theoretical studies employing
density-functional theory (DFT) were carried out in order to
identify these barriers on the potential-energy surface
(PES)~\cite{honkala00,yourdshahyan01,yourdshahyan02}, but were
unable to find any sizeable barriers. Consequently, although no
explicit calculation of the sticking curve based on the
high-dimensional PES was done, it was concluded that the adiabatic
PES studied by these DFT calculations cannot explain the
experimental findings.

This has led to speculations that non-adiabatic effects may play an
important role in the oxygen dissociation process at the Al(111)
surface~\cite{kasemo74,kasemo79,yourdshahyan01,katz04,wodtke04,hellman03,hellman05},
just as much as for the interaction of O$_2$ with several other
metal
surfaces~\cite{boettcher90,kasemo74,greber94,zhdanov97,kato00,kato98}.
In an earlier letter we investigated this role of non-adiabatic
effects in the oxygen dissociation at Al(111) within a
first-principles approach~\cite{behler05}, and traced the
non-adiabatic effects back to spin-selection rules~\cite{wigner27}
conserving the initial spin-triplet character of the free O$_2$
molecule during the molecule's approach to the surface. In the
present paper, we now give a detailed account of our approach and
its underlying multi-faceted methodology. This methodology comprises
the calculation of spin-constrained PESs within our recently
introduced locally-constrained DFT (LC-DFT) method \cite{behler06},
and a neural network interpolation to obtain the PESs in closed form
for the six dimensions representing the molecular degrees of
freedom. Ensuing extensive molecular dynamics (MD) simulations on
the continuous PESs are then used to obtain the sticking probability
as the proper statistical average over many trajectories with
different initial molecular orientations and positions over the
surface unit-cell.

With this methodology, we first firmly establish that an adiabatic
description based on the ground state Born-Oppenheimer
PES~\cite{born27} can indeed not explain the experimentally reported
low sticking probability of thermal molecules. Emphasizing the
importance of spin-selection rules in the molecule-surface
interaction, we model non-adiabatic effects by confining the
trajectories of the impinging O$_2$ molecules to motion on a PES, in
which the spin-triplet character of the gas-phase molecule is
conserved. This leads indeed to a significant reduction of the
sticking probability at lower kinetic energies, in qualitative
agreement with experiment. In order to further substantiate these
findings of our previous work~\cite{behler05}, we also critically
discuss major uncertainties of our approach, which arise from
confining the O$_2$ trajectories exclusively on the spin-triplet
constrained PES, and from the employed approximate
exchange-correlation (xc) functionals.

While this analysis reveals noticeable quantitative effects, it also
 clearly shows that spin-selection rules are indeed the ruling factor
 behind the reduced sticking probability for thermal molecules. Due to
 the hindered spin transition the largest fraction of the impinging
 molecules is already repelled into the gas phase at rather large
 distances from the surface where other mechanisms like charge transfer
 from the metal are still quite weak. Only for the remaining molecules
 such alternative mechanisms (and corresponding other electronic states)
 will start to play a role, and will then most likely lead to dissociation.

\section{The Role of Spin-selection Rules for the Oxygen Dissociation at Al(111)}\label{section2}

When investigating the role of non-adiabatic effects, it is crucial
to first define, what the adiabatic ground state of the system is
and to what kind of non-adiabaticity one refers to. Central for the
determination of the sticking probability are trajectories of
impinging molecules, which start at a very large distance from the
surface, and, in the case of a successful dissociation event, end
with two individual oxygen atoms. For large molecule-surface
separations the ground state of the system is given by an O$_2$
molecule in its $^3\Sigma_g^-$ triplet ground state and a
non-magnetic Al(111) surface, whereas adsorbed at the surface the
oxygen adatoms are as non-magnetic as the Al(111) surface. Sometime
during the molecular trajectory the initial spin moment on the
oxygen is therefore quenched, and in an adiabatic calculation this
happens gradually during the molecule's approach to the surface. In
fact, we will see below that in adiabatic DFT calculations this
quenching occurs already partly when the molecule is still
(infinitely) far apart, where wave functions do not yet overlap.

Interestingly, for an isolated O$_2$ molecule such a spin-transition
during dissociation would be forbidden by Wigner's spin-selection
rules \cite{wigner27}, which state that in any elementary step of
bond making or breaking, the total spin of the reactants and
products must be the same. In other words, the spin of the reactants
must be conserved, or transferred to some other entity. In the case
of the O$_2$ dissociation at the Al(111) surface, this other entity
to which the spin is transferred is the solid surface. However, if
this transfer probability is still low at larger distances from the
surface, the spin-selection rules imply that the probability for an
electronic transition toward the O$_2$ spin-singlet state
($^1\Delta_g$) will be small, even if the surface potential shifts
the $^1\Delta_g$ energy below the spin-triplet $^3\Sigma_g^-$
energy. Small probability means in this context a probability of
similar magnitude as the spin-transfer probability in the gas-phase,
where the lifetime of spin-singlet O$_2$ is 72 minutes. Due to this
low transition probability, the central assumption of the adiabatic
picture may then become incorrect, which is that the electrons are
at any time able to instantaneously follow the motion of the nuclei
and remain in their electronic ground state.

Instead of molecular trajectories on the adiabatic PES, molecules
conserving their initial spin-triplet character during the approach
to the surface are then the relevant ones for the surface scattering
and the determination of the sticking coefficient. This situation
can be conveniently described in a diabatic-like picture, where the
electronic structure at the O$_2$ molecule is constrained to a
certain spin state.\cite{behler06} Relevant non-adiabatic states of
the interacting system correspond then to the ground and certain
excited states of the system, when the O$_2$ molecule is still so
far away from the surface that there is no interaction. This would
be a state corresponding to the O$_2$ molecule in its $^3\Sigma_g^-$
spin-triplet gas-phase ground state and the Al(111) surface in its
singlet ground state (henceforth simply termed spin-triplet state).
Another state corresponds to the O$_2$ molecule in its $^1\Delta_g$
spin-singlet state and the Al(111) surface in its singlet ground
state (henceforth termed spin-singlet state). Further constrained
states include situations where electrons are transferred from the
metal surface to the O$_2$ molecule, e.g., the case of an O$_2^-$
ion and an Al(111)$^+$ surface (henceforth termed ionic state).

\begin{figure}[t!]
\scalebox{0.55}{\includegraphics{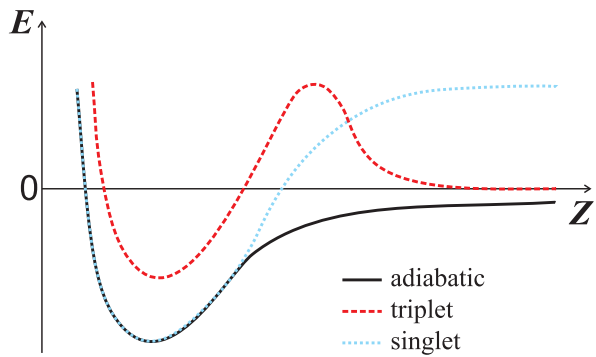}} \caption{(Color online)
Schematic energy diagram showing the two spin constrained
potential-energy surfaces (PESs) and the adiabatic PES as a function
of the molecule-surface separation $Z$. The adiabatic ground state
PES (solid line) has the lowest energy for all $Z$. Molecules travel
initially on the spin-triplet PES (dashed line) and if this PES
exhibits an energy barrier as shown here, lower energy molecules
will be reflected, leading to a corresponding reduction in the
sticking probability. The spin-singlet PES (dotted line) is an
excited state at large $Z$, but becomes lower in energy for small
$Z$.} \label{model}
\end{figure}

In this non-adiabatic picture the system dynamics is given by motion
on and transitions between the various PESs corresponding to the
different constrained states. At the beginning of the molecular
trajectories, i.e., the largest molecule-surface distances, the
energetic order of these PESs follows the excitation spectrum of the
free O$_2$ molecule: 
The spin-triplet state is lowest in energy,
followed by the spin-singlet state, which in experiment is about
1~eV higher in energy. The next state of interest
is the ionic state which is higher than the spin-triplet state by
the difference of the Al(111) work function and the O$_2$ electron
affinity. Experimentally this energy difference is about 4~eV. 
The molecular
trajectory starts in the spin-triplet state. At decreasing
molecule-surface distances, the various constrained states are
influenced differently by the surface potential, and eventually, the
spin-singlet state will result as the lowest energy state.

If we focus for the moment on only the two energetically lowest
constrained states at large distances, the situation is thus as
shown schematically in Fig. \ref{model}. Since the adiabatic picture
is not spin-consistent, the adiabatic PES is lowered by a coupling
of the spin-triplet and spin-singlet states, which is largest at the
crossing point of the two spin-constrained PESs. Due to this
lowering, the adiabatic PES can therefore be barrierless, even if
the spin-triplet PES exhibits a barrier to dissociation as shown in
Fig. \ref{model}. In the adiabatic picture, the molecule would then
dissociate along with a gradual quenching of its spin moment.
However, if the crossing of the spin-triplet and spin-singlet PES
happens at distances from the surface that are large enough to
prevent an efficient spin transfer to the Al(111) surface, the
spin-selection rules will prevent the transition from the
spin-triplet to the spin-singlet state and the molecular motion will
continue on the spin-triplet PES up to distances to the surface,
where the increased coupling to the surface allows an efficient
spin-transfer. If the triplet PES then exhibits barriers as shown
schematically in Fig. \ref{model}, low energy molecules would not be
able to overcome these barriers and would be reflected back into the
gas-phase, explaining the experimentally measured low sticking
coefficient for thermal O$_2$ molecules.

Central for a validation of this conceptual understanding is
therefore the computation of the spin-triplet PES surface. Important
is not only whether this PES indeed exhibits barriers to
dissociation, but also at which molecule-surface distances these
barriers are. The closer the barriers are located to the surface,
the higher is the coupling to the surface and the lower the
probability that the molecular motion will remain on the
spin-triplet PES. In this respect it is important to note that at
small molecule-surface distances, also a multitude of other
constrained states including the above discussed ionic state will
become energetically favorable. While Fig. \ref{model} focuses on
only the two spin-states for simplicity, it is then in principle the
coupling and electronic transitions to all these states that will
play a role. However, if the spin-triplet PES exhibits barriers at
molecule-surface distances, where the energetic order of the PESs
remains still largely as dictated by the gas-phase excitation
spectrum, then only the spin-triplet state and at most transitions
to the spin-singlet state are relevant for the determination of the
sticking coefficient. A crucial first step is therefore to
calculate, based on first principles, the spin-triplet PES and
identify the size and location of possible energy barriers, as well
as to analyze their implications for the sticking curve.

\section{Methodology}\label{methodology}

\subsection{Calculation of the Sticking Curves}\label{stickingcurve}

The experimentally measured sticking coefficient is an average over
a large number of molecule-surface collisions with different initial
molecular configurations. These configurations comprise different
molecular orientations with respect to the surface, as well as
different lateral positions of the molecule's center-of-mass over
the surface unit-cell. A statistically reliable sticking curve can
therefore only be calculated from a larger number of MD trajectories
with random initial configurations that account for these degrees of
freedom. We address this challenge by employing a divide and conquer
approach.
In this approach the calculation
of the sticking curve is split into three consecutive steps~\cite{gross95,gross98a}.
In a
first step, the PES of the molecule-surface interaction is mapped as
a function of the relevant coordinates. In the second step, this
precalculated mesh is then interpolated or fitted by a continuous
representation. Such an interpolation of high-dimensional PESs is a
tedious task, but in recent years, several techniques, like
analytical fits~\cite{gross95,wiesenekker96,dai97,wei98,kluner98},
tight binding representations~\cite{gross99,gross03,mehl96,cohen94},
genetic programming~\cite{makarov98} and the modified Shephard
method~\cite{bettens99,crespos03}, possibly combined with a
corrugation-reduction scheme~\cite{busnengo00,kresse00}, have been
developed. In the present work we employ a very general neural
network fitting scheme, which has already proven to be a powerful
tool for the accurate representation of multi-dimensional PESs in
similar applications~\cite{blank95,brown96,lorenz04,lorenz06}. Since
the evaluation of the energy and forces from the neural network
representation is about 5 to 6 orders of magnitude faster than
direct {\itshape ab initio} calculations, a large number of MD
trajectories can be calculated in the last step of the ``divide and
conquer'' approach to obtain the sticking probabilities at various
molecular kinetic energies.

\subsection{Calculation of the Potential-Energy Surfaces}

\begin{table}
\begin{ruledtabular}
\begin{tabular}{lll}
Functional            & $E_{\rm b}$ (eV) & $r$ (\AA) \\ \hline
LDA~\cite{perdew92a}  & 7.43            & 1.21 \\
PBE~\cite{perdew96}   & 6.10            & 1.22 \\
PW91~\cite{perdew92b} & 6.07            & 1.22 \\
BLYP~\cite{gill92}    & 5.75            & 1.24 \\
RPBE~\cite{hammer99}  & 5.65            & 1.23 \\ \hline
Experiment            & 5.12 \cite{herzberg52} & 1.21~\cite{herzberg50} \\
\end{tabular}
\end{ruledtabular}
\caption{Binding energies, $E_{\rm b}$, and bond lengths, $r$, of
the free O$_2$ molecule in its $^3\Sigma_g^-$ ground state, as
obtained with our present computational setup and different
exchange-correlation functionals. The zero point energy of about
0.1~eV has not been taken into account, but is part of the
experimental results.} \label{o2properties}
\end{table}

All DFT calculations have been carried out with the full-potential
all-electron code DMol$^3$ using numerical atomic orbitals as basis
functions~\cite{delley90,delley00}. For the oxygen atoms the
so-called {\itshape all} basis is used (19 atomic orbitals per
atom), for the aluminium atoms the {\itshape dnd} basis is applied
(18 atomic orbitals per atom). These basis sets are essentially
equivalent to a ``double numeric'' basis, i.e., the valence orbitals
are described by a linear combination of atomic orbitals of the free
atom and of the free positive ion, which is further improved by a
set of polarization functions. For details of the basis set we refer
to Ref.~\onlinecite{delley90}. A real-space cutoff of 9 bohr has
been applied to the basis functions. The irreducible wedge of the
first Brillouin zone is sampled by 10 k-points. To improve
convergence a Pulay mixing scheme~\cite{pulay80,pulay82} has been
employed combined with a thermal Fermi broadening of 0.1~eV,
extrapolating the energies to zero temperature~\cite{wagner98}. To
check on the dependence of the obtained results on the
exchange-correlation functional, the full PESs have been calculated
using two different functionals, the PBE~\cite{perdew96} and the
RPBE~\cite{hammer99} functional. Even though both functionals are
based on the generalized gradient approximation (GGA) and provide a
much better description of the oxygen molecule than the
local-density approximation~\cite{perdew92a} (LDA), they yield
rather different binding energies of the free oxygen molecule as
shown in Table~\ref{o2properties}. In this respect, we use both
functionals to obtain a first idea about the uncertainties in our
results with respect to the description of electronic exchange and
correlation. Compared to experiment all functionals included in
Table~\ref{o2properties} yield a strong overestimation of the
binding energy of the oxygen molecule. The reason for this
overbinding has been described in detail by Gunnarsson and
Jones~\cite{gunnarsson85} and is due to an insufficient error
compensation between the oxygen atom and the molecule, as the
different nodal structures of the wave functions of these two
species are not fully taken into account in the description of the
exchange energy in jellium based exchange-correlation functionals
(e.g. LDA and GGAs). The consequences of this error in the molecular
binding energy on our study of the dissociation process will be
discussed in more detail in Section~\ref{adiapes}.

The Al(111) surface is modelled by a (3$\times$3) slab geometry
consisting of 7 aluminium layers. Tests with different supercells
have shown that for the often employed (2$\times$2) supercells the
interaction energy between the oxygen molecules in neighboring cells
is still about 0.3~eV for an extended, but in the present context
still relevant molecular bond length of 2.4~{\AA} in a parallel
orientation to the surface. In the $(3 \times 3)$ cells, for
comparison, this interaction is reduced to about 70~meV for this
particular bond length. Since the clean Al(111) surface shows only a
marginal outward relaxation of about 1~\% of the interlayer
distance~\cite{jacobsen95,fall98,kiejna01}, all atoms have been
fixed in their bulk positions. The aluminium lattice constants
obtained with PBE and RPBE are very similar (4.05~{\AA} and
4.08~{\AA}), and the PBE lattice constant has therefore been used in
all calculations. To avoid interactions between the periodically
repeated slabs a large vacuum of 30~{\AA} has been used. This does
not increase the computational effort due to the finite range of the
localized basis functions, but enables us to perform calculations
for molecule-surface separations of up to 10~\AA. Making use of
inversion symmetry oxygen is adsorbed at both sides of the slab to
prevent dipole interactions between the slabs through the vacuum
region. At the largest molecule-surface separations considered in this work, the O$_2$ molecule is then still more than 10~\AA{} away from its periodic image impinging on the other side of the slab.

\begin{figure}[t!]
\scalebox{0.45}{\includegraphics{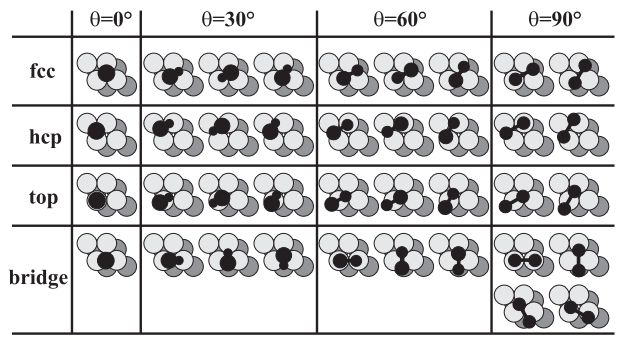}} \caption{Top views of the
molecular configurations for which elbow plots have been calculated
in the mapping of the six-dimensional potential-energy surface. For
clarity only four of the nine surface Al atoms in the (3$\times$3)
supercell are shown (grey spheres). The two oxygen atoms are shown
as small black circles and the relative size of both atoms indicates
their vertical height (larger sphere means further away from the
surface and closer to the viewer). Correspondingly, only one atom is
visible in configurations, in which the molecule approaches in a
vertical orientation. $\theta$ is the angle between the surface
normal and the molecular axis.} \label{geos}
\end{figure}

\begin{figure}[t!]
\scalebox{1.85}{\includegraphics{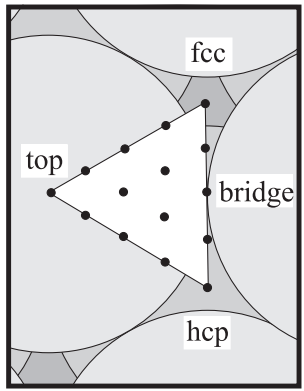}} \caption{Top view explaining
the symmetry of the Al(111) surface. The irreducible wedge spanned
by the fcc, hcp, and top sites is shown as white triangle. The
surface atoms are represented by the grey circles. The surface sites
within the irreducible part of the unit-cell used for the mapping of
the barrier region of the spin-triplet potential-energy surface are
shown as black dots.} \label{wedge}
\end{figure}

When keeping the nuclei in the Al(111) surface at fixed positions,
the PES for the oxygen dissociation is still six-dimensional due to
the molecular degrees of freedom. The commonly employed coordinate
system is a superposition of a Cartesian coordinate system for the
center-of-mass of the oxygen molecule ($X$, $Y$, and $Z$, the latter
being the direction perpendicular to the surface) and a spherical
coordinate system for the oxygen-oxygen bond length $r$, the angle
between the molecular axis and the surface normal $\theta$, and the
angle $\phi$ between the projection of the molecular axis in the
$xy$-plane and the positive $x$-axis. The six-dimensional PES is
mapped by calculating a number of two-dimensional cuts, in which the
energy is given as a function of $r$ and $Z$ and which are commonly
called ``elbow plots'' due to their characteristic shape. All other
degrees of freedom, i.e., $\theta$, $\phi$, $X$ and $Y$ are fixed in
a single elbow plot, and the configuration space referring to these
four dimensions is mapped by calculating elbow plots for many
different surface sites and many different molecular orientations.
In the case of the adiabatic PES, a total of 38 different elbow
plots have been calculated for each xc functional and the molecular
configurations are detailed in Fig.~\ref{geos}. The symmetry of the
surface has been fully exploited in that only energies for
configurations within the irreducible wedge of the unit-cell spanned
by the fcc, hcp and top site have been calculated, as explained in
Fig.~\ref{wedge}. The energy zero point has been defined as the
total energy for an infinite separation between the surface and the
molecule, with the latter at its equilibrium bond length. This bond
length of the free O$_2$ molecule as obtained with the different
functionals is also compiled in Table~\ref{o2properties}.

As will be apparent below, the spin-triplet PES exhibits barriers to
dissociation, which depend strongly on the molecular orientation
with respect to the surface. Since also the height of the barriers
depends strongly on the molecular orientation towards the surface, a
more detailed mapping of the barrier region than in the adiabatic
case was done to ensure an accurate description of the PES for the
MD simulations. In addition to the full elbow plots at the
high-symmetry fcc, hcp, top, and bridge sites used for the adiabatic
PES, cf. Fig. \ref{geos}, configurations with $r$=1.3~{\AA} and
$Z$=2.1~{\AA} close to the maximum of the barrier have been
calculated for all possible angular orientations of the molecule at
the 11 ``off-symmetry'' sites shown in Fig.~\ref{wedge} by sampling
the angles $\theta$ and $\phi$ in 30$^\circ$ intervals.
Consequently, the six-dimensional shape of the energy barrier is
mapped very accurately, which allows to properly account for
steering effects, if present, in the molecular dynamics simulations.

\subsection{Locally-Constrained Density Functional Theory}\label{constraint}

While the adiabatic ground state PES is accessible by standard
(spin-polarized) DFT calculations, the calculation of
spin-constrained PESs requires the introduction of appropriate
constraints on the electronic configurations. In order to localize
the triplet spin at the oxygen molecule for all molecule-surface
separations we developed and employed a locally-constrained DFT
technique in the spirit of the seminal work of Dederichs {\itshape et
al.}~\cite{dederichs84}, which is also
very similar to the independently developed approach by
Wu and Van Voorhis~\cite{wu05}. For a detailed description of this
technique and a comparison to other approaches we refer to
Ref.~\onlinecite{behler06} and present here only a concise summary.

The central idea in the LC-DFT technique employed in this work is to
split the system into two subsystems, the oxygen molecule and the
Al(111) surface, and to constrain the electron numbers of these
subsystems according to the constrained state of interest. In the
general case the calculations are spin-polarized, and for the two
subsystems four electron numbers $N_{\rm O_2}^{\uparrow}$, $N_{\rm
O_2}^{\downarrow}$, $N_{\rm Al}^{\uparrow}$, and $N_{\rm
Al}^{\downarrow}$ have to be considered. These electron numbers are
determined by projecting the Kohn-Sham states into the Hilbert
spaces of the two subsystems, which in the case of a localized
atomic orbital basis set are spanned by the basis functions centered
at the atoms of the respective subsystems. This projection yields
the partial densities of states (pDOS) of the four subsystems, which
are then filled separately with the four electron numbers
corresponding to the non-adiabatic state of interest. In the general
case, this yields four different Fermi energies $\epsilon_{\rm
F,O_2}^{\uparrow}$, $\epsilon_{\rm F,O_2}^{\downarrow}$,
$\epsilon_{\rm F,Al}^{\uparrow}$, and $\epsilon_{\rm
F,Al}^{\downarrow}$, which are subsequently aligned under the
constraint of fixed electron numbers. This alignment is achieved by
the introduction of a configuration-dependent auxiliary potential
$H_{\rm A}^{\sigma}$. $H_{\rm A}^{\sigma}$ is a projection operator
into the oxygen subspace and thus allows to shift $\epsilon_{\rm
F,O_2}^{\sigma}$ with respect to $\epsilon_{\rm F,Al}^{\sigma}$. The
auxiliary potential is added to the standard DFT Hamiltonian
$H_0^{\sigma}$ to yield a new effective Hamiltonian $H^{\sigma}_{\rm
eff}$ for each spin $\sigma$
\begin{equation}
H^{\sigma}_{\rm eff} \;=\; H_{0}^{\sigma} \;+\; H_{\rm A}^{\sigma} \quad .
\label{hamilt}
\end{equation}
The strength of $H_{\rm A}^{\sigma}$ is adjusted self-consistently
until the Fermi energies are aligned, i.e., the electronic structure
problem is solved fully self-consistently {\em under the constraint
of fixed electron numbers in the two subsystems and for each spin
channel}.

The electronic structure of the system is thus completely relaxed
under the given constraint (formulated as an additional external
potential) and in this sense, the LC-DFT method calculates a
``constrained electronic ground state'' based on the Hohenberg-Kohn
theorem~\cite{hohenberg64}. The method is very general, and by
adapting the electron numbers in principle arbitrary constrained
PESs can be calculated for different spin and charge states of the
subsystems. We stress, however, that this constraint does, of
course, not allow to overcome limitations related to the employed
exchange-correlation functional (which, e.g., affects the multiplet
structure~\cite{jones89} of some systems) and we will come back to
this point below.

\subsection{Neural Network Interpolation}

In this paper we closely follow the neural network (NN) fitting
procedure described in detail in Refs.~\onlinecite{lorenz04},
\onlinecite{lorenz06} and summarize here only the features that will
be relevant for the later sections. We only note that a different
way of incorporating the symmetry of the surface into the NN has
been used which no longer contains any approximations and which is
described in detail elsewhere~\cite{behler07}. Basically, this
procedure consists of a coordinate transformation of the original
six molecular coordinates to a set of symmetry functions, 11 in this
case, that is used as input for the NN.

In general, a neural network is a non-linear technique that allows
to fit any function to a high accuracy and does not require any
knowledge about the functional form of the underlying
problem.~\cite{hornik89} In particular multilayer feed-forward
neural networks have already successfully been employed to provide
accurate fits of potential-energy
surfaces~\cite{blank95,lorenz04,lorenz06,brown96}. In the present
work this NN type is applied to fit a smooth and continuous function
to the DFT data by optimizing a set of parameters. This function has
an analytic form and consequently the forces required in the MD
simulations can easily be obtained from the derivatives of this
function. This way the energies and forces are consistent, which is
essential for MD applications.

\begin{figure}[t!]
\scalebox{0.6}{\includegraphics{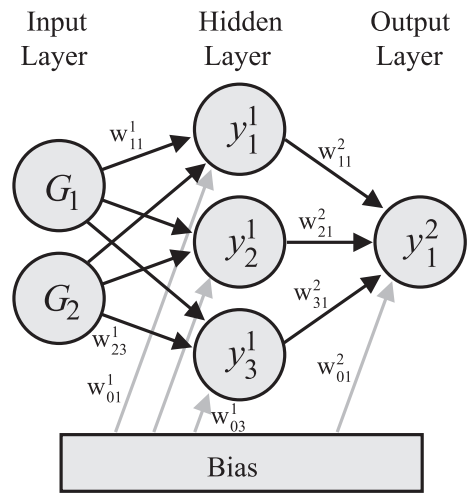}} \caption{Structure of a
\{2-3-1\} feed-forward neural network (NN). The gray circles
represent the nodes of the network. Each node $j$ in layer $k$ is
connected to the nodes $i$ in the preceding layer by the weights
$w_{ij}^k$, the parameters of the NN. The bias adds a constant to
each node in the hidden and output layers to adapt the positions of
the non-linearity intervals of the activation functions. The output
$y_1^2$ is the potential energy $E$ and is calculated according to
Eq.~(\ref{nnequation}).} \label{nnscheme}
\end{figure}

The general form of a small feed-forward NN is shown schematically
in Fig.~\ref{nnscheme}. It consists of several layers, each of which
contains one or more nodes represented by the grey circles. In the
given example two nodes in the input layer represent two arbitrary
coordinates $G_1$ and $G_2$ determining the total energy $E$. This
energy corresponds to the single node in the output layer. Between
the input and the output layer one hidden layer with three nodes is
located. The term hidden layer is used, because the numerical values
at the nodes of this layer are just auxiliary quantities without a
physical meaning. Each node $i$ of a certain layer is connected with
the node $j$ in the subsequent layer via a weight parameter
$w_{ij}^k$, where $k$ represents the index of the target layer. On
each node all the values from the nodes of the preceding layer are
summed after being multiplied by the connecting weight. On the
resulting sum a non-linear function $f_{\rm a}$ is applied. This
function is called activation function and is typically a
sigmoidally shaped function that introduces the non-linearity
capability to the NN. Specifically, we use the hyperbolic tangent as
activation function in this work. For very large or very small
arguments the activation functions converge to a constant number,
but for a certain interval the output changes in a non-linear way
allowing to fit very complex functions. The bias unit acts as an
adjustable offset to adapt the position of the non-linearity
interval of the activation functions. Then the obtained function
value of each node is passed to all the nodes of the subsequent
layer and multiplied by the corresponding connecting weights. In the
output layer the values are collected and the output value is
calculated by applying a linear function as activation function. In
a feed-forward NN information is transferred only in one direction
through the network, from the input layer via the hidden layer(s) to
the output layer. In general, each layer including input and output
layers can contain many more nodes than shown in the simple example
in Fig.~\ref{nnscheme}, and also more than one hidden layer is
typically used.

With this construction a complicated non-linear function relating
the input coordinates to the energy is defined. This function can be
given analytically and it depends on many parameters, the weights
$w_{ij}^k$. For the example given in Fig.~\ref{nnscheme}, this
function is
\begin{equation}
\label{nnequation}
E\left(\mathbf{G}\right) \;=\; f_{\rm a}^2 \left( w_{01}^2 + \sum_{j=1}^{3}
w_{j1}^2 f_{\rm a}^1 \left(w_{0j}^1 + \sum_{i=1}^{2} w_{ij}^1 G_i
\right) \right) \quad,
\end{equation}
where $y_{i}^k$ labels the value of node $i$ in layer $k$, and for
the input nodes we have used the symbol $\mathbf{G}=\{G_i\}$.
$w_{0j}^k$ is the bias weight for the activation function $f_{\rm
a}^k$ acting on node $j$. In this work we use a short hand notation
to describe the structure of a neural network. The example in
Fig.~\ref{nnscheme} is a \{2-3-1-tl\} network, i.e., 2 nodes in the
input layer, 3 nodes in one hidden layer, and 1 node in the output
layer. ``t'' indicates the use of a hyperbolic tangent as activation
function in the hidden layer, while a linear ``l'' function has been
used in the output layer.

In order to calculate the output, i.e., the interpolated energy, the
connecting weight parameters have to be known. They are determined
by training the NN using a set of known DFT data points. An
independent test set of DFT data, which has not been used in the
parameter optimization is used to check the accuracy of the fit for
geometries not included in the training set. For the adiabatic PES
computed with the PBE functional 1723 DFT energies were calculated
and split into a training and a test set. The test set contains 79
randomly chosen points and is not used for the optimization of the
network parameters. After extensive testing the best fit was
achieved with a \{11-38-38-1-ttl\} NN structure. The mean average
deviations (MAD) with respect to the original DFT data are 38~meV
and 62~meV in the training and test set, respectively. The root mean
squared error (RMSE) of the full training set is 0.080~eV, for the
test set it is 0.105~eV. To obtain a good fit particularly in the
region of the PES that is accessible in the MD runs, i.e., the
regions with a potential energy of less than +1~eV with respect to
the sum of the isolated subsystems, in all fits reported in this
work gradually higher fitting weights have been assigned to points
close to the minimum energy path. This reduces the RMSE along the
minimum energy path in the entrance channels of the elbow plots to a
few meV and reduces also significantly the maximum deviation we
obtain for the data points that are fitted worst. While this maximum
fitting error is with 0.69 eV still sizeable for the data set with
energies up to +1 eV, only 18 out of the 704 data points that
correspond to the entrance channel of an impinging thermal O$_2$
molecule in a MD simulation exhibit errors that are larger than
50~meV. We carefully checked these few single data points with high
fitting error to make sure that they do not introduce artificial
topological features to the PES or affect the sticking curve
simulations (which are most sensitive to the PES barriers). A
similarly good fit has been achieved for the adiabatic PES computed
with the RPBE functional. Here we calculated 1369 DFT energies, and
the test set consisted of 69 randomly chosen points. The training
and test MADs obtained with a \{11-40-40-1-ttl\} network are 22~meV
and 50~meV, the RMSE are 0.041~eV and 0.100~eV, respectively. For
both xc functionals all calculated elbow plots are perfectly
reproduced by the NN fits.

Due to the more complicated, barrier-containing topology of the
spin-triplet PES, we used a larger number of DFT data points for the
NN interpolation. Specifically, 2870 and 2917 DFT energy points were
computed for the RPBE and the PBE xc functional, respectively. For
the interpolation 96 of the RPBE energies and 97 of the PBE energies
were chosen randomly as test data set. The best fit for the RPBE
data points is obtained from a \{11-40-40-1-ttl\} NN and has a MAD
of 0.023 and 0.033~eV for the training and test points,
respectively. The corresponding RMSEs are 0.049~eV and 0.070~eV. In
case of the PBE PES the MADs are 0.035 and 0.031~eV, the RMSEs
0.078~eV and 0.050~eV, respectively, obtained also with a
\{11-40-40-1-ttl\} NN. Again, we found the biggest error in a single
data point in potential energy regions up to +1~eV to be still
relatively large (0.43~eV for RPBE), but due to the employed fitting
weights, this largest error in the data points that are accessible
by a thermal O$_2$ molecule is reduced to 17 meV in the case of the
RPBE PES. With the ultimately achieved fits, all 38 elbow plots, but
also the additional points at the off-symmetry sites are accurately
reproduced. Further details of the fitting procedure and an error
analysis are published elsewhere~\cite{behler07}.

\subsection{Molecular Dynamics Simulations}

The calculation of a statistically reliable sticking curve requires
the calculation of molecular trajectories with different initial
configurations (initial lateral position and orientation of the
molecule with respect to the surface). The MD simulations are based
on solving Hamilton's equations of motion,
\begin{eqnarray}
\dot{q}_i &=&  \frac{\partial H}{\partial p_i} \\
\dot{p}_i &=& -\frac{\partial H}{\partial q_i} \quad,
\end{eqnarray}
with the classical nuclear Hamiltonian
\begin{eqnarray}
H=\frac{1}{2M_{\rm O_2}}\left(p_{X}^2+p_{Y}^2+p_{Z}^2\right)+ \nonumber \\
\frac{1}{2\mu}\left(p_r^2+\frac{p_{\theta}^2}{r^2}+\frac{p_{\phi}^2}{r^2\sin^2\theta}\right)+V\left(\mathbf{R,r}\right) \quad ,
\end{eqnarray}
where $p$ is the momentum, $M_{\rm O_2}$ the mass and $\mu$ the
reduced mass of the oxygen molecule. $V\left(\mathbf{R,r}\right)$ is
our NN representation of the potential energy. For the numerical
integration, we employ a Burlisch Stoer
algorithm~\cite{numericalrecipes} with a variable time step to
improve the efficiency and accuracy.

Initially, the oxygen molecule has the equilibrium gas-phase bond
length and is placed 9.5~\AA{} above the surface. The angular
orientation of the molecule and the lateral position are chosen
randomly. The initial velocity of the molecule towards the surface
is determined by the translational kinetic energy and the lateral
velocity components are zero resulting in a perpendicular angle of
incidence of the molecule as in experiment~\cite{osterlund97}. The
trajectories are assumed to yield a dissociation event, if the O$_2$
bond length is stretched beyond 2.4 {\AA}, i.e., the bond length has
doubled, and the momentum $p_{r}$ is still positive. 
We have verified that the specifics of this dissociation criterion plays no role for the sticking curves discussed in this work. Specifically the lowering of the triplet sticking curve results from those molecules directly repelled at the triplet PES barriers described below. Molecules that overcome these barriers will predominantly dissociate, and a discussion of the specific mechanisms behind this is outside the scope of the present work.
The molecule is
considered as reflected, if $Z$ exceeds 6~{\AA} with a positive
$p_{Z}$, i.e., the molecule is leaving the surface. The sticking
curve is determined for kinetic energies ranging from 0.025~eV to
0.975~eV in steps of 50~meV. From tests for the spin-triplet
constrained case (see below) we found that the qualitative form of
the sticking curve was already obtained, when averaging over around
20 trajectories with random initial O$_2$ orientation for each
kinetic energy. The error in the sticking probability $S$ for $N$
 trajectories is given by $\sqrt{\frac{S(1-S)}{N}}$. Since the MD
 runs on the interpolated NN-PESs are computationally inexpensive,
 the fully converged sticking curves presented here were obtained
 by averaging over 2000 trajectories for each kinetic energy.

\section{Results}

\subsection{The Adiabatic Picture}\label{adiapes}
\subsubsection{The Adiabatic Potential-Energy Surface}

\begin{figure*}[t!]
   \epsfig{bbllx=0,bblly=0,bburx=587,bbury=433,clip=,
           file=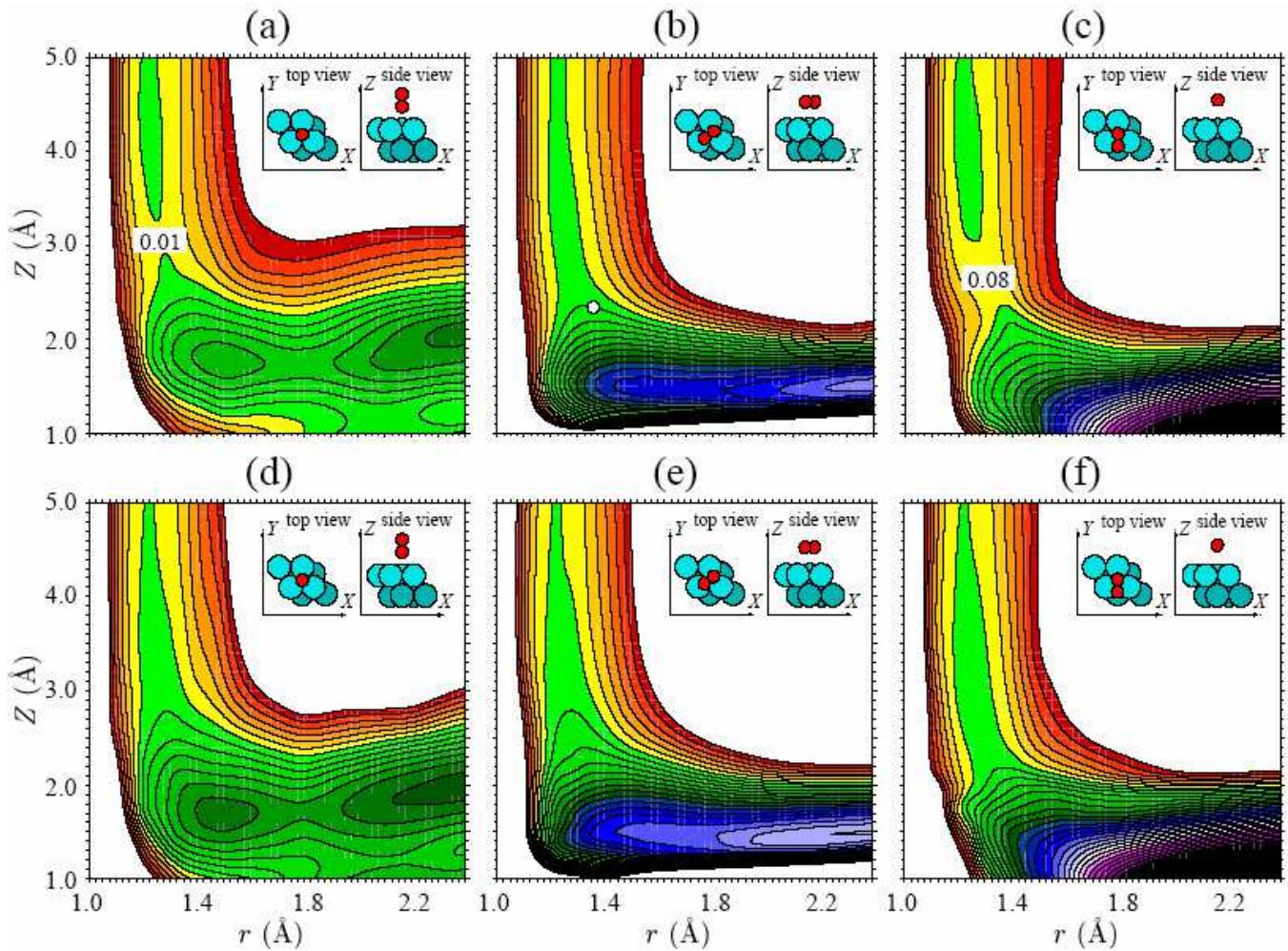,width=1.0\linewidth}
\caption{(Color online)
Two-dimensional cuts (``elbow plots'') through the six-dimensional
adiabatic potential-energy surface (PES) for oxygen dissociation at
the Al(111) surface. The energy is shown as a function of the
center-of-mass distance of the molecule from the surface $Z$ and the
oxygen-oxygen bond length $r$. In (a), (b) and (c) the elbow plots
obtained with the RPBE functional are shown for the three different
molecular orientations explained in the insets. In (d), (e) and (f)
the corresponding elbow plots obtained from the PBE functional are
shown. Energy barriers are labelled in eV. The energy difference
between the contour lines is 0.2~eV.\label{adiabatic}}
\end{figure*}

We begin our study with the adiabatic PES for the oxygen
dissociation at Al(111). Three of the calculated elbow plots are
shown in Fig.~\ref{adiabatic} for both the RPBE and the PBE
functional. In case of the RPBE functional we find several elbow
plots with shallow energy barriers of up to 0.1~eV, while in case of
the PBE functional none of the calculated elbow plots shows a
barrier towards dissociative adsorption. These results are in good
agreement with previous calculations of parts of this PES in smaller
surface unit-cells~\cite{honkala00,yourdshahyan01,yourdshahyan02}.
We only mention in passing that there is a general problem when
calculating the adiabatic PES for the dissociation of oxygen at
Al(111) with standard DFT implementations. There is a charge
transfer from the metal to the oxygen molecule that is present even
for an infinite molecule-surface separation. This is because the
antibonding 2$\pi^{\ast\downarrow}$ O$_2$ orbital, which is
unoccupied in the free molecule, is lower in energy than the Fermi
level of the metal surface. Consequently, electron density is
transferred from the surface to the molecule, which raises the
energy of the 2$\pi^{\ast\downarrow}$ orbital until it is aligned
with the Fermi level. This charge transfer to the molecule is of the
order of 0.01 $e$ for large, even macroscopic distances from the
surface as has been discussed in detail in
Ref.~\onlinecite{behler06}.

A direct comparison of the elbow plots obtained with the RPBE and
the PBE functional shows that the RPBE PES is less attractive, which
is a typical feature of this functional and is consistent with
previous adsorption studies~\cite{hammer99}. However, the energy
difference between the PBE and the RPBE PES is far smaller than the
0.5~eV binding energy difference of the free O$_2$ molecule obtained
with these functionals, cf. Table \ref{o2properties}. This indicates
that the strong overbinding of the O$_2$ molecule does not carry
through to the shape of the adiabatic PES. The latter is instead
determined by energy differences of different molecular
configurations, which are likely to have similar errors in the xc
energy.

When the oxygen molecule approaches the surface, the energetically
more favorable orientation is initially perpendicular to the surface
in agreement with previous
studies~\cite{yourdshahyan01,yourdshahyan02}. Closer to the surface
the parallel configuration is preferred, which allows a stronger
interaction of both atoms with the surface. During the approach to
the surface, there is a continuous charge transfer from the surface
to the molecule, and a continuous reduction of the triplet spin
until finally the singlet state of the adsorbed adatoms is adopted.
The oxygen dissociation is a strongly exothermic process and the
final binding energy of an oxygen atom with respect to a free O$_2$
molecule is about 4.1~eV (PBE) and 3.8~eV (RPBE) in the employed
(3$\times$3) supercell. At higher coverages this even increases due
to attractive oxygen-oxygen interactions~\cite{kiejna01}, i.e., from
an energetic point of view, both abstraction and dissociative
chemisorption of both atoms are possible.

\subsubsection{The Adiabatic Sticking Curve}

\begin{figure}[t!]
\scalebox{0.3}{\includegraphics{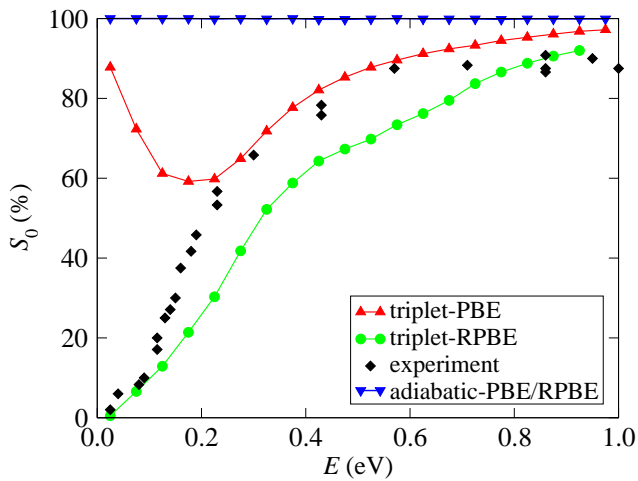}} \caption{(Color online)
Sticking curves obtained from MD simulations on the adiabatic and
the spin-triplet potential-energy surfaces as calculated with two
different exchange-correlation functionals (PBE and RPBE). The
experimental data (solid diamonds) are taken from
Ref.~\onlinecite{osterlund97}. $E$ is the translational kinetic
energy of the oxygen molecules.} \label{tripletsticking}
\end{figure}

Due to the almost complete absence of energy barriers in the
computed adiabatic PES (for both xc functionals) a high sticking
probability is very likely to be found. Nevertheless, slow molecules
might be dynamically steered to certain surface sites and angular
orientations which can have a strong effect on the sticking
properties~\cite{gross95}. It is not possible to assess such effects
by a mere inspection of low-dimensional cuts of the PES, and we
correspondingly carried out extensive MD simulations on the
interpolated PES. For both xc functionals, the sticking probability
was found to be basically unity independent of the kinetic energy,
which is in strong disagreement with experiment as shown in
Fig.~\ref{tripletsticking}. In case of the PBE functional this is
immediately obvious due to the complete absence of energy barriers
towards dissociation. Yet, also the shallow energy barriers found
for some elbow plots calculated with the RPBE functional do not
reduce the sticking probability. This has two reasons. First, the
height of these barriers is very small, so that they can be overcome
even by thermal molecules in most cases. Second, slow molecules,
which are most likely to be stopped by the barriers, are steered~\cite{gross98} 
towards the barrier-free entrance channels.

Careful tests to determine the dependence of the sticking curves on
the actual NN fit and the number of MD trajectories have been
carried out, all with the same result, the sticking probability is
essentially 100~\% for thermal molecules. The identical result
obtained with two xc functionals that yield a largely different
O$_2$ binding energetics, cf. Table \ref{o2properties}, renders it
somewhat unlikely that the accuracy of the employed GGA functionals
is insufficient for the description of the dissociative adsorption
process.  Finally, we also performed 24 on-the-fly {\em ab initio} MD runs for thermal
O$_2$ molecules, where the full dynamics of the Al atoms was taken
into account. In all cases the O$_2$ molecule dissociated and the
bond length was already notably weakened at molecule-surface
distances, where significant O-induced movement of the Al atoms had
not even set in. This suggests that neither the rigid substrate
approximation underlying our approach may be held responsible for
the dramatic disagreement with the experimental data. We therefore
have to conclude that the low sticking probability of thermal oxygen
molecules impinging on the Al(111) surface cannot be explained by
calculations based on the adiabatic PES obtained from
state-of-the-art DFT.

\subsection{The Non-Adiabatic Picture}\label{diabaticpes}
\subsubsection{The Spin-Triplet Potential-Energy Surface}\label{fsm}

\begin{figure}[t!]
\scalebox{0.5}{\includegraphics{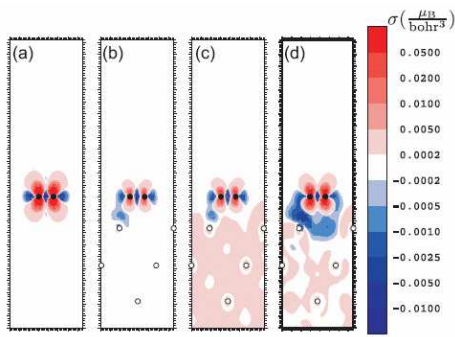}} \caption{(Color online)
Magnetization-density plot obtained with the RPBE functional and for
a molecule with a bond length of 1.3~{\AA}, oriented parallel to the
surface and 2.1~{\AA} above the fcc site. This molecular
configuration corresponds to a position on the triplet energy
barrier (indicated by the white circle in Fig.~\ref{adiabatic}(b)
and Fig.~\ref{tripletpes}(b)). Shown is a side view, with the upper
half corresponding to the vacuum and the lower half to the Al
surface (white circles represent the positions of Al atoms in the
shown cut plane, and black circles the positions of the two O
atoms). In (a) the magnetization density of a free O$_2$ molecule in
its triplet ground state without the aluminium slab is shown. (b)
gives the reduced magnetization-density in the adiabatic
calculation. In (c) the magnetization-density distribution for a
triplet fixed spin moment calculation is shown, while (d) refers to
a triplet constrained calculation.}\label{magnetization}
\end{figure}

\begin{figure*}[t!]
\scalebox{1.0}{\includegraphics{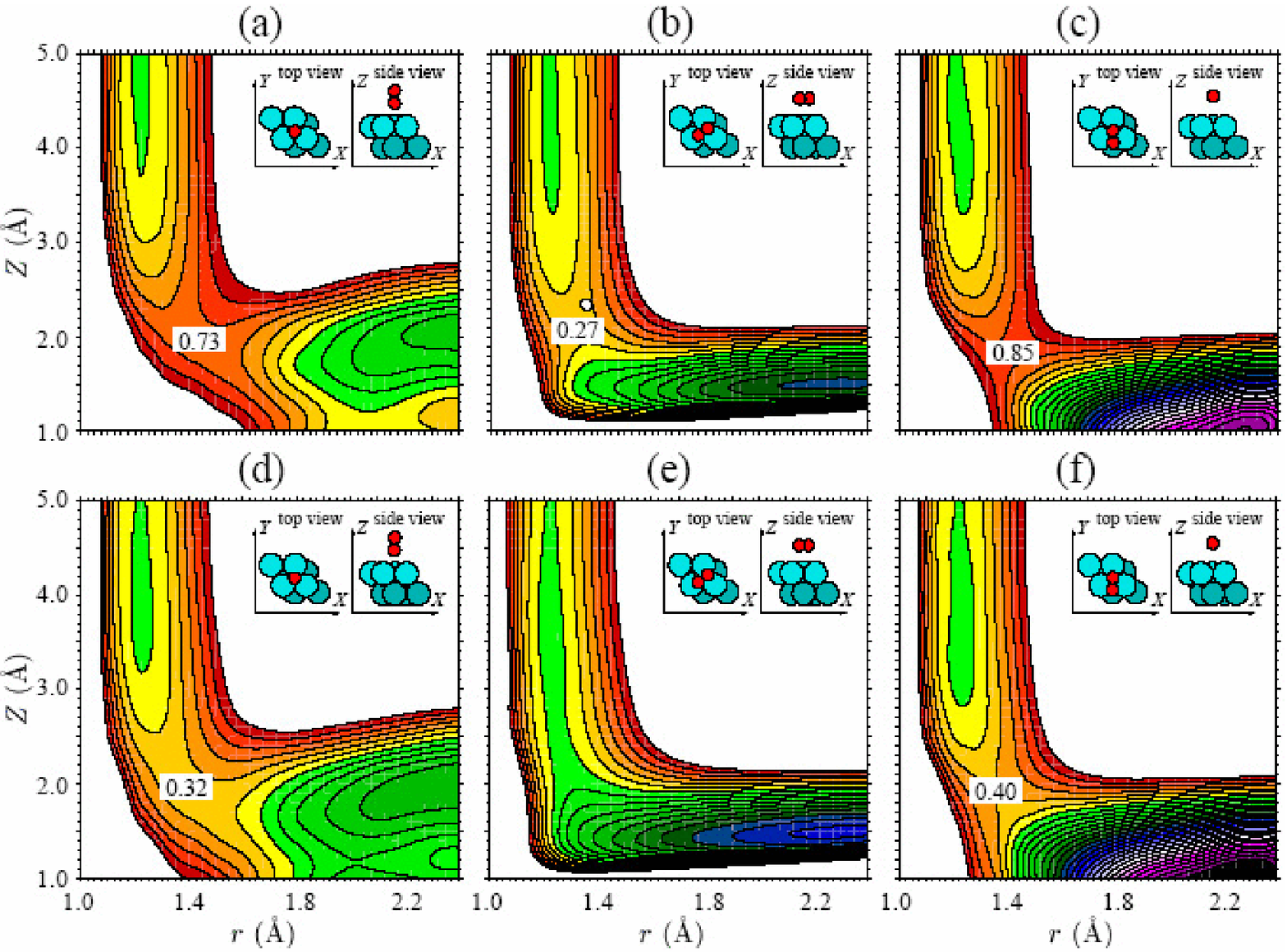}} \caption{(Color online)
Two-dimensional cuts (``elbow plots'') through the six-dimensional
spin-triplet potential-energy surface (PES) for the oxygen
dissociation at the Al(111) surface obtained with the
locally-constrained DFT approach. Equivalent to Fig. \ref{adiabatic}
the energy is shown as a function of the center-of-mass distance of
the molecule from the surface $Z$ and the oxygen-oxygen bond length
$r$. In (a), (b) and (c) the elbow plots obtained with the RPBE
functional are shown for the three different molecular orientations
explained in the insets. In (d), (e) and (f) the corresponding elbow
plots obtained from the PBE functional are shown. Energy barriers
are noted in eV. The energy difference between the contour lines is
0.2~eV.\label{tripletpes}}
\end{figure*}

If we therefore turn to the role of non-adiabatic effects, the
discussion in Section \ref{section2} suggests to focus on the
question of whether the spin-triplet constrained PES exhibits
dissociation barriers and at which height over the surface they are
located. A straightforward approach to this crucial PES would be to
constrain the overall system to a spin-triplet configuration by
means of the fixed spin moment (FSM)~\cite{schwarz84} technique.
Here, no constraint on the spatial distribution of the magnetization
density is made, and only the overall numbers of spin-up and
spin-down electrons in the system are fixed. We computed the PES
with this approach, but obtained barrierless PESs that very much
resemble the adiabatic ones. The reason is that in the FSM
calculations close to the surface the triplet spin is becoming more
and more delocalized. For Z~$<$~1.8~\AA{} the triplet spin is in
fact almost completely delocalized in the interior of the aluminium
slab and almost no magnetization density is left on the oxygen
molecule as illustrated in Fig.~\ref{magnetization}. Since this is
not the physics we want to describe, a different approach to the
spin-triplet PES is required that locally confines the magnetization
density to the oxygen molecule.

We achieve this confinement with the LC-DFT approach~\cite{behler06}
that allows to control the electron numbers in the oxygen and the
aluminium subsystems for each spin explicitly. Using this method the
spin-triplet PES has been calculated employing the PBE and the RPBE
functional~\cite{functional}. In Fig.~\ref{tripletpes} three of the
38 calculated elbow plots are shown for both functionals. In
contrast to the adiabatic PES sizeable energy barriers exist with
heights up to 0.9~eV.~\cite{comparison} In the RPBE case there are barriers for all
configurations, the lowest being about 0.05~eV for a molecule
oriented parallel to the surface above a bridge site with $\phi$=0°
(not shown). The typical stronger binding of the PBE functional
reduces the barriers in the PES computed with this functional by
about 0.3-0.4~eV compared to the RPBE case. For 6 of the 38
calculated elbows, the barrier disappears therefore even completely
(all for parallel molecular orientations). Such a pronounced
functional-dependence has not been found in the adiabatic PES, and
we will see below the consequences this has on the calculated
sticking curve. Apparently, the exchange part, which is the only
difference between the PBE and the RPBE functional~\cite{hammer99}
is particularly sensitive to the triplet configuration close to the
surface, a region where most or all of the spin has vanished in the
adiabatic calculations.

In order to understand the origin of the energy barrier it is
instructive to analyze the magnetization density at the energy
barrier. This is shown in Fig.~\ref{magnetization} for a molecule
parallel to the surface above an fcc site for the RPBE functional.
In (a) the magnetization density of a free oxygen molecule in its
spin-triplet ground state without the metal surface is shown for
comparison. In (b) the strongly reduced magnetization density of an
adiabatic calculation is given. The integrated adiabatic
magnetization density amounts to 0.064 $\mu_{\rm B}$ instead of 2.0
$\mu_{\rm B}$ for the ideal triplet. The magnetization density in
(c) corresponds to the triplet FSM calculation. Although the
integrated magnetization density still corresponds to a full
triplet, a large amount of the spin has been transferred to the
aluminium slab, which is the reason why the FSM method does not
yield energy barriers. In (d) the magnetization density for the
triplet LC-DFT calculation is plotted. The triplet spin is localized
at the oxygen molecule causing a depletion of spin-up density in the
surrounding region of the metal surface because of the Pauli
repulsion of like spins. This increased Pauli repulsion is thus the
origin of the energy barrier. As the metal surface is in an overall
singlet state, the displaced aluminium spin-up density is
delocalized in the slab, but in contrast to the FSM calculation no
net magnetization of the slab is present.

\subsubsection{The Spin-Triplet Sticking Curve}\label{tripletmd}

MD simulations have been carried out on the spin-triplet PES for
both xc functionals. These simulations are complementary to the MD
trajectories on the adiabatic PES, in that the latter represent the
case of an instantaneous charge transfer and spin reduction when the
molecule approaches the surface, while for the MD confined to the
triplet PES no charge transfer and no change in the spin is allowed
at any time. This corresponds therefore to the limit of an infinite
lifetime of the molecule on the spin-triplet PES, which is a good
approximation if the motion of the molecule is fast compared to the
probability for electronic transitions. The resulting sticking
curves are shown in Fig.~\ref{tripletsticking}. The sticking curve
based on the PBE triplet PES is very similar to the RPBE curve for
medium and high kinetic energies. The upshift of the PBE sticking
coefficient with respect to the RPBE values can be explained by the
systematically lower energy barriers in case of the PBE functional
and provides a kind of ``functional error bar''. For low kinetic
energies, however, the PBE sticking curve deviates qualitatively
from the RPBE curve and the experimental data. The reason for the
high sticking probability of about 87~\% for thermal molecules in
this case is the absence of energy barriers for a few molecular
configurations. Thermal molecules approach the surface very slowly
and have sufficient time to adapt their orientation and lateral
position to the shape of the surface potential. The majority of the
lowest energy molecules is therefore steered into the few
barrier-free entrance channels and can adsorb at the surface,
whereas this dynamical steering becomes less efficient with
increasing kinetic energy of the molecules.

The difference in the sticking probabilities of thermal molecules
obtained with the two xc functionals shows how critical smallest
energy differences in the PES can be for the resulting sticking
curve. Although the PBE triplet PES is clearly less attractive than
the adiabatic PES, in the typical barrier region the energy of the
few barrierless configurations is only about 20~meV lower than the
vacuum level. Very small energy increases in this region would
therefore be sufficient to yield energy barriers. In fact, if the
energy were only by about 50~meV higher, both functionals would
yield qualitatively the same sticking curve. Although there is thus
a pronounced xc functional dependence, a significant reduction of
the sticking probability is nevertheless obtained for both
functionals. This reduction is not present in the adiabatic case, no
matter which functional is used.

\begin{figure}[t!]
\scalebox{0.35}{\includegraphics{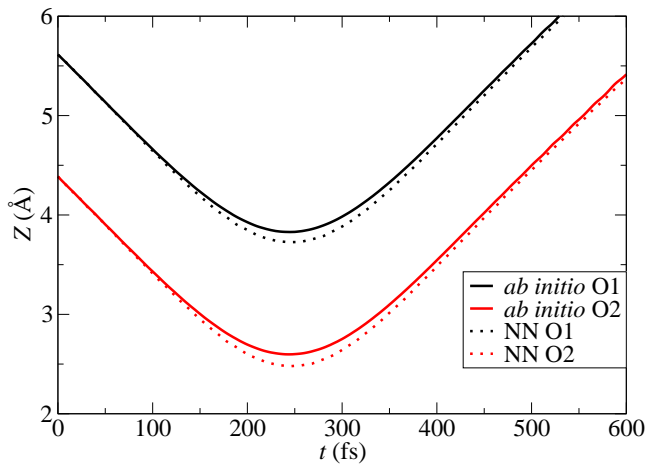}} \caption{(Color online)
Comparison of a molecular trajectory on the spin-triplet PES
obtained once with the NN-interpolated PES employing the
frozen-surface approximation (solid line) and once with a direct
{\em ab initio} MD run allowing full motion of the Al atoms (dashed
line). The molecule is initially located above the fcc site and
oriented perpendicular to the surface with an initial kinetic energy
of 0.15~eV. Shown is the vertical distance of the O atoms from the
rigid first layer Al atoms as a function of time $t$. The atom
closer to the surface is labelled with ``O2'', the atom further away
from the surface with ``O1''. Due to the onset of the energy
barriers the turning point of the molecule is far away from the
surface. No significant coupling to the motion of the surface atoms
occurs and the NN and {\itshape ab initio} trajectories are
basically identical.\label{nnabinitio}}
\end{figure}

Finally, we again employed on-the-fly {\em ab initio} MD simulations
to check on the validity of the rigid substrate approximation
underlying our approach. In Fig.~\ref{nnabinitio} the vertical
positions of the two oxygen atoms in the course of a trajectory are
shown for a molecule oriented perpendicular to the surface above an
fcc site. The dotted lines represent the positions of the oxygen
atoms as obtained from the NN-based MD employing the frozen-surface
approximation. The molecule has an initial kinetic energy of
0.15~eV, which is not sufficient to overcome the barrier, and
consequently it is reflected back into the gas-phase. The straight
black lines represent the oxygen positions in a corresponding direct
{\itshape ab initio} MD run for the same initial atomic positions.
The trajectory of the molecule is hardly changed. The induced motion
of the aluminium atoms is very small and sets in only once the
molecule has approached the surface to about 3\,{\AA}. We conclude,
that in the present system the frozen-surface approximation is
well-justified for reflected molecules. If,
however, a molecule has a high enough energy to overcome the
barrier, it necessarily dissociates and the mobility of the surface
Al atoms is then crucial in determining the further trajectory.

\section{Discussion}\label{discussion}

The results presented in the last Section can be summarized in that
the experimental low sticking probability for thermal molecules can
not be explained within an adiabatic picture. When confining the
motion of the incoming molecules to the spin-triplet PES, there is
some variation depending on the employed xc functional, but overall a
significant lowering of the sticking coefficient is obtained. Of
course, the latter represents only the opposite to the adiabatic
picture: In the adiabatic picture the electronic structure adapts
instantaneously to the nuclear motion, in the spin-triplet motion
the electronic structure remains always in the initial gas-phase
spin-state. In reality transitions away from this spin-triplet PES
to other constrained states may occur. However, before embarking on
this, it is necessary to assess, how well the employed xc functional
may actually describe these various states, which we discuss here
first for the isolated O$_2$ molecule.

\subsection{O$_2$ multiplets}

It is well known that in DFT the description of multiplets can be
substantially incorrect~\cite{jones89}. This is because the central
quantity in DFT is the electron density, while the nodal structure
of the wave function is not explicitly taken into account. The tiny
differences in the electron (and magnetization) densities of
non-degenerate multiplets are not resolved in present-day LDA or GGA
xc functionals, so that these multiplets turn out as degenerate.
Similarly, degenerate multiplets that do differ in the electron (and
magnetization) densities may be recognized as non-degenerate. This
problem has been analyzed by Gunnarsson and Jones for a series of
molecular
states~\cite{gunnarsson77,jones89,gunnarsson85,gunnarsson80},
showing that the origin of this limitation lies in the construction
of the electron density from a single determinant of Kohn-Sham
orbitals. This typically yields an accurate energy with current GGA
functionals, if the wave function of the system can be represented
correctly by a single determinant. If, however, the wave function is
a combination of two or more determinants, the symmetry information
of the state is not taken into account in present-day xc
functionals~\cite{ziegler77,barth79,gunnarsson80}.

\begin{figure}[t!]
\scalebox{0.5}{\includegraphics{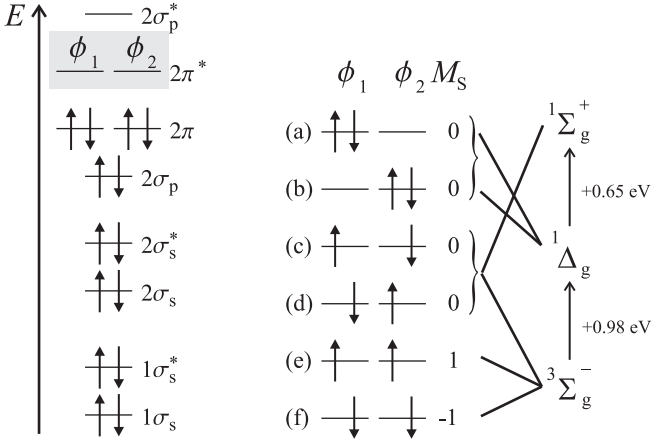}} \caption{Molecular orbital
diagram of the O$_2$ molecule. The $^3\Sigma_g^-$ ground state and
the two excited $^1\Delta_g$ and $^1\Sigma_g^+$ singlet states
differ only in the occupation numbers of the doubly degenerate
$2\pi^{\ast}$ state shown in gray on the left side. The 6 possible
occupations are shown in (a)-(f) on the right side. Since open shell
determinants are no eigenfunctions of the spin operator unless all
electrons have parallel spin, 4 of the 6 oxygen wave functions are
linear combinations of the determinants (a)-(d). Additionally shown
are the measured energy gaps between the triplet and singlet
states~\cite{herzberg50}.}\label{o2terms}
\end{figure}

The oxygen molecule is a di-radical, i.e., the highest occupied
molecular orbital, the antibonding $2\pi^{\ast}$ orbital, is
two-fold degenerate and occupied by two electrons.\cite{salem72} The
low-lying singlet and triplet states of the O$_2$ molecule, which
differ only in the occupation of this orbital, are given by the wave
functions
\begin{subequations}\label{dets}
\begin{eqnarray}
\psi_1^{\rm s} &=& \frac{1}{2} \bigg[\phi_1\phi_1 - \phi_2\phi_2\bigg] (\alpha\beta -\beta\alpha) \\
\psi_2^{\rm s} &=& \frac{1}{2} \bigg[\phi_1\phi_1 + \phi_2\phi_2\bigg] (\alpha\beta -\beta\alpha) \\
\psi_3^{\rm s} &=& \frac{1}{2} \bigg[\phi_1\phi_2 + \phi_2\phi_1\bigg] (\alpha\beta -\beta\alpha) \\
\psi_4^{\rm t} &=& \frac{1}{2} \bigg[\phi_1\phi_2 - \phi_2\phi_1\bigg] (\alpha\beta +\beta\alpha) \\
\psi_5^{\rm t} &=& \frac{1}{\sqrt{2}} \bigg[\phi_1\phi_2 - \phi_2\phi_1\bigg] (\alpha\alpha) \label{trip1}\\
\psi_6^{\rm t} &=& \frac{1}{\sqrt{2}} \bigg[\phi_1\phi_2
-\phi_2\phi_1\bigg] (\beta\beta) \quad .\label{trip2}
\end{eqnarray}
\end{subequations}
Here, we have used for the involved spatial orbitals the short-hand
notation $\phi_1\phi_1=\phi_1(1)\phi_1(2)$ and for the spin
functions $\alpha\beta=\alpha(1)\beta(2)$, omitting the electronic
labels. There are thus three degenerate triplet (t) $^3\Sigma_g^-$
states ($\psi_4^{\rm t},\psi_5^{\rm t},\psi_6^{\rm t}$), a two-fold
degenerate singlet (s) $^1\Delta_g$ state ($\psi_1^{\rm
s},\psi_2^{\rm s}$), and a higher singlet $^1\Sigma_g^+$ state
($\psi_3^{\rm s}$). The experimental singlet-triplet gap between the
$^3\Sigma_g^-$ and the $^1\Delta_g$ state for the free molecule is
about 0.98~eV, and between the $^3\Sigma_g^-$ and the $^1\Sigma_g^+$
state the energy difference is 1.63~eV~\cite{herzberg50}. The energy
diagram of the oxygen states is shown schematically in
Fig.~\ref{o2terms}. The triplet states shown in Fig.~\ref{o2terms}
(e) and (f) are represented by single determinants ("high-spin"
triplets) and we may consequently expect that Kohn-Sham DFT with LDA
or GGA functionals is in principle able to properly describe these
triplet states. In a standard spin-polarized calculation they are
found as the ground state of O$_2$, and, correspondingly, we are
confident that the calculated spin-triplet constrained PES correctly
describes the initial state in the dissociation process at the
Al(111) surface.

For the singlet states, which like the triplet state are
di-radicals, the situation is different. None of the singlet wave
functions can be represented by a single determinant and it is thus
{\itshape a priori} not clear if these states are described
sufficiently accurately by Kohn-Sham DFT with the present-day xc
functionals. For the singlet-triplet gap, i.e., the energy
difference by which the singlet state is higher than the triplet
state, we compute half of the experimental value (0.392~eV for PBE,
0.393~eV for RPBE). The reason is that a symmetry-broken electron
density is obtained corresponding to one of the electronic
configurations (c) and (d) in Fig.~\ref{o2terms} with a non-zero
magnetization density. This is a common observation in Kohn-Sham
DFT~\cite{ovchinnikov96,sorescu01}. In the same way as $\psi_3^{\rm
s}$ and $\psi_4^{\rm t}$ are linear combinations of the two Slater
determinants (c) and (d) in Fig.~\ref{o2terms}, the single Slater
determinant corresponding to the electronic configuration that we
obtained is a combination of the ``low-spin'' triplet $^3\Sigma_g^-$
and the $^1\Sigma_g^+$ state, which is commonly referred to as
``spin contamination''~\cite{hehre86}. Because of the large error in
the computed singlet-triplet gap and the wrong spin-up and down
densities we conclude that Kohn-Sham DFT with current xc functionals
does not allow to calculate the singlet state with sufficient
accuracy. An alternative without proper formal justification is
to calculate a singlet state simply using non-spinpolarized DFT
calculations. In this case the magnetization density is zero
everywhere in space. The calculations are straightforward and yield
a singlet triplet gap of 1.138~eV (PBE) and 1.171~eV (RPBE) for the
free O$_2$ molecule, i.e., much closer to the experimental value.

Yet another possibility to obtain the singlet-PES would be an application
of the ``sum method'' of Ziegler, Rauk and Baerends~\cite{ziegler77}. In this
method the energy of the singlet state is calculated from the single-determinant singlet energy and the energy of the triplet state. This approach, as well as the non-spinpolarized approach involve the mapping of a singlet PES and are thus expected to lead to similar computational costs. For the free O$_2$ molecule, also the results are essentially the same. Since the relevant part of the singlet-PES for the sticking coefficient is the one located at rather large distances from the surface, i.e., the part where the singlet-PES is still quite gas-phase like, we therefore expect that the sum method and the approach based on non-spinpolarized calculations that will be employed here yield essentially the same results. 

\subsection{Effect of electronic transitions on the sticking curve}

The barriers found on the spin-triplet PES are located at rather
large distances from the surface. As discussed in Section
\ref{section2}, the energetic order of the various constrained
states is then still predominantly governed by the gas-phase O$_2$
excitation spectrum. With the ionic states still very high
in energy (and thereby ruling out the charge transfer picture suggested
by Hellman et al.~\cite{hellman05}), the only alternative constrained state in
closer energetic vicinity to the spin-triplet state is thus the
spin-singlet state, which will be the first alternative constrained
state to become lower in energy than the spin-triplet state at
decreasing molecule-surface distances as schematically shown in Fig.
\ref{model}. Motivated by Wigner's spin-selection rules and because
the interaction time is not very long (cf. Fig.~\ref{nnabinitio}),
we had disregarded such transitions to the spin-singlet state in the
MD simulations confined to the spin-triplet PES. We now set out to
scrutinize this assumption.

With decreasing molecule-surface distance, the coupling to the solid
surface increases and thereby the spin-selection rules become
weakened. Crucial for the analysis of possible electronic
transitions from the spin-triplet to the spin-singlet state is
therefore to determine the location of the high-dimensional crossing
seam of the two corresponding PESs and to compare this position to
the location of the energy barriers on the spin-triplet PES. If the
crossing is at distances from the surface larger than the onset of
the barriers on the spin-triplet PES, electronic transitions to the
spin-singlet state are likely to modify the result obtained for the
sticking coefficient in the spin-triplet only MD simulations. If the
crossing is at distances closer to the surface than the onset of
barriers, low energy molecules will be reflected by the barriers on
the spin-triplet PES before ever approaching the surface close
enough for electronic transitions to play a role. In this case, the
 result for the sticking curve obtained with the spin-triplet MD
simulations will be quite accurate.

Central for such an analysis is thus the knowledge of both the
spin-triplet  and the spin-singlet PES. Specifically, it is only the parts
of these PESs at the larger distances from the surface where the
triplet barriers are located that are of relevance for the here discussed sticking
coefficient. As discussed above, there
are unfortunately strong indications that present-day xc functionals
may not be able to describe the spin-singlet PES with sufficient
accuracy. Motivated by the closeness of the calculated
singlet-triplet gap for the free O$_2$ molecule to the experimental
value \cite{chargetransfer}, we therefore choose to use
non-spinpolarized calculations to obtain a spin-singlet PES, even
though there is no formal justification that this should indeed
yield a good approximation to the real spin-singlet state. A weak
rationalization for this choice may be that with the correct singlet-triplet
splitting at large distances from the surface, the right energetic
displacement between the two PESs should be provided. If then the
approximate spin-singlet PES has roughly the same shape as the real
spin-singlet PES, this should be good enough to track the
approximate location of the crossing seam between the two PESs,
which is at present our only interest.

\begin{figure*}[t!]
\scalebox{0.9}{\includegraphics{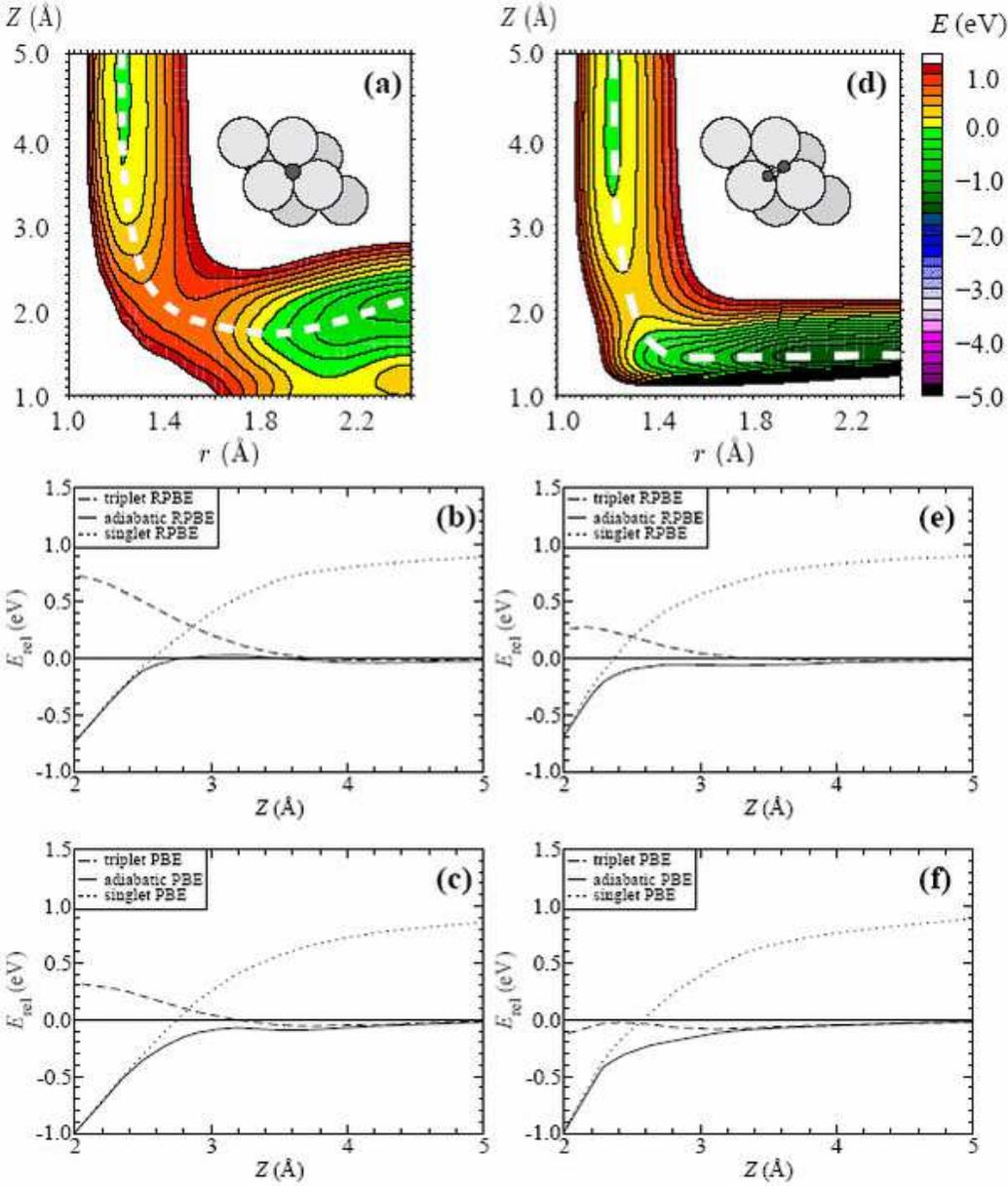}} \caption{(Color online)
Energetics along the reaction path for a molecule impinging above
the fcc site, with (a) a perpendicular and (d) a parallel
orientation to the surface. The white dashed line corresponds to the
minimum energy path for this molecular configuration. In (b) and (c)
the energies of the adiabatic PES (black line), the approximate
singlet PES (dotted line) and the LC-DFT spin-triplet PES (dashed
line) along this path are shown as a function of the center-of-mass
distance from the surface $Z$. (b) corresponds to the RPBE
functional, (c) refers to the PBE functional. The corresponding PESs
for the parallel orientation are shown in (e) for the RPBE and in
(f) for the PBE functional.}\label{reactionpath}
\end{figure*}

In Fig.~\ref{reactionpath} the minimum energy path for dissociation
on the spin-triplet PES is shown for two different molecular
orientations. Along this minimum energy path, the energetics is
additionally shown for the two xc functionals for the adiabatic, the
LC-DFT spin-triplet and the non-spinpolarized approximate
spin-singlet case. It can clearly be seen that the crossing of the
spin-triplet and the spin-singlet PES is for both molecular
orientations closer to the surface than the onset of the energy
barriers on the spin-triplet PES. A similar result is obtained for
molecules impinging at several other sites over the surface.
Low-energy, thermal molecules are therefore most likely reflected by
the energy barriers on the spin-triplet PES before they can reach
distances to the surface, where electronic transitions would set in
with a higher probability. For the lower kinetic energies, we
therefore do not expect notable changes in the sticking curve due to
electronic transitions compared to the result obtained with the
spin-triplet confined MD simulations. Molecules with a higher
kinetic energy, on the other hand, may overcome the barrier and
reach the crossing seam of the two PESs. Electronic transitions may
then occur with a higher probability. However, molecules with a high
kinetic energy of about 1~eV do even dissociate on the spin-triplet
PES, and thus possible electronic transitions will not be able to
increase the sticking coefficient much further compared to the
anyway close to 100\% value obtained by the spin-triplet only MD
simulations. At best, electronic transitions may therefore only have
a notable effect on the sticking curve at intermediate kinetic
energies, most likely further increasing the sticking probability.
In this respect it is intriguing to note that the spin-triplet
 confined sticking coefficient computed with the RPBE functional
 in Fig.~\ref{tripletsticking} does indeed lie below the 
experimental curve at these
 intermediate energies. However, in light of the xc functional 
dependence discussed in Section IV.B.2 it would be unwarranted
 to ascribe this difference directly to the effect of electronic
 transitions.

\section{Conclusion}\label{summary}

In the present paper we have further detailed our first-principles
investigation of the initial dissociative sticking of oxygen
molecules at the Al(111) surface. We provide evidence that
fundamental and undisputable experimental results like the low
sticking probability for thermal oxygen molecules cannot be
explained in terms of an adiabatic dissociation process. In order to
show this we calculated the adiabatic potential-energy surface in
six-dimensions using density-functional theory. Subsequent
interpolation by a neural network fitting procedure and extensive
molecular dynamics simulations on the interpolated PES shows
unambiguously that the adiabatic description yields a unit sticking
probability independent of the molecular kinetic energy, in clear
disagreement with the experimental data.

As most likely non-adiabatic effects cause this discrepancy, we
focused on a hindered spin-flip of the incoming O$_2$ molecule away
from the initial spin-triplet gas-phase state. For this purpose we
employed a new locally-constrained DFT method that allows to
localize the triplet spin at the oxygen molecule, and were thereby
able to compute the corresponding spin-triplet PES seen by such a
molecule when impinging on the surface. In contrast to the adiabatic
PES, this PES exhibits barriers in the molecular entrance channel.
Consequently, we find a reduced sticking probability when confining
the molecular motion to this spin-triplet PES. 
While we can argue that electronic transitions away from this PES
to other states will have basically no effect on thermal molecules
and will only slightly increase the sticking probability for higher
kinetic energies, we do observe
 a substantial
variation in the calculated sticking probability depending on the
employed xc functional. Future studies employing a higher-level
energetics are therefore desirable for a final quantitative
comparison to the experimental data.

\section{Acknowledgements}

The authors gratefully acknowledge Bernard Delley for providing the
DMol$^3$ code, into which we implemented our LC-DFT approach. We
wish to thank Eckart Hasselbrink and Bengt Lundqvist for stimulating
discussions.


\begin{thebibliography}{99}
\bibitem{functional}
In case of the spin-triplet PES only the PBE energies have been calculated fully self-consistently. Starting from the self-consistent PBE energies, the RPBE energies have been obtained from one additional iteration employing this functional. We have carefully checked that the error introduced by this non-self-consistent procedure is negligible by comparing the results of 96 fully self-consistent RPBE energies with the corresponding non-self-consistent values. The average error in energy differences between different molecular configurations was found to be about 1.3~meV, while the maximum error found was 6~meV.
\bibitem{chargetransfer}
When calculating the energetic difference between the spin-triplet PES and the approximate spin-singlet PES obtained by the non-spinpolarized calculations, we obtain 0.898~eV and 0.937~eV for the PBE and RPBE xc functional at large molecule-surface distances (6\,{\AA}), respectively. These values differ from those obtained for the isolated O$_2$ molecule by about 0.2\,eV. The reason is again the uncontrolled charge transfer to the oxygen $2\pi^{\ast}$ orbital that has been discussed in detail in Ref. \onlinecite{behler06}. This charge transfer artificially lowers the total energy of the non-spinpolarized singlet PES, by coincidence such that almost exactly the experimental value is obtained for the singlet-triplet gap.
\end{thebibliography}
\end{document}